\documentclass[aps,prl,reprint,groupedaddress]{revtex4-1}

\usepackage{graphicx}
\usepackage{dcolumn}
\usepackage{bm}
\usepackage{amsmath,amssymb,latexsym}

\begin{document}

\newcommand\lh{L_h}
\newcommand\lb{L_e}
\newcommand\lp{L_p}
\newcommand\lbl{L_{b}}
\newcommand\rd{\textrm{d}}
\def\strutdepth{\dp\strutbox}
\def\nw#1{\strut\vadjust{\kern-\strutdepth\vtop to0pt{\vss\hbox to\hsize
{\hskip\hsize\hskip5pt$\leftarrow$\hss\strut}}}{\em #1}}


\title{Viscously Controlled Peeling of an Elastic Sheet by Bending and Pulling}

\author{John R. Lister}
\author{Gunnar G. Peng}
\affiliation{Institute of Theoretical Geophysics, Department of Applied Mathematics and Theoretical Physics, University of Cambridge, Wilberforce Road, Cambridge, CB3 0WA, UK}

\author{Jerome A. Neufeld}
\affiliation{BP Institute, Department of Earth Sciences, Department of Applied Mathematics and Theoretical Physics, University of Cambridge, Bullard Laboratories, Madingley Road, Cambridge, CB3 0EZ, UK}


\date{\today}
\begin{abstract}
Propagation of a viscous fluid beneath an elastic sheet is controlled by local dynamics at the peeling front, in close analogy with the capillary-driven spreading of drops over a precursor film. Here we identify propagation laws for a generic elastic peeling problem in the distinct limits of peeling by bending and peeling by pulling, and apply our results to the radial spread of a fluid blister over a thin pre-wetting film.  For the case of small deformations relative to the sheet thickness, peeling is driven by bending, leading to radial growth as $t^{7/22}$.  Experimental results reproduce both the spreading behaviour and the bending wave at the front.  For large deformations  relative to the sheet thickness, stretching of the blister cap and the consequent tension can drive peeling either by bending or by pulling at the front, both leading to radial growth as $t^{3/8}$.  In this regime, detailed predictions give excellent agreement and explanation of previous experimental measurements of spread in the pulling regime in an elastic Hele-Shaw cell~\cite{puzovic-2012}.
\end{abstract}             
             


\maketitle

The viscous spreading of fluid beneath an elastic sheet is controlled by dynamics at the peeling front, in close analogy to the control exerted by the contact line in the capillary spreading of drops over a precursor film.  Dynamical control of fluid-mediated elastic peeling can be found in, for example, the manufacture of flexible electronics and MEMS~\cite{hosoi-2004,rogers-2010}, the reopening of airways~\cite{jensen-2002,grotberg-2004}, the suppression of viscous fingering in a deformable Hele-Shaw cell~\cite{puzovic-2012,al-housseiny-2013}, and the geological formation of laccoliths~\cite{michaut-2011,bunger-2011}  by the lateral flow of lava beneath an elastic sediment layer.  
 
The controlling influence of contact lines in the related problem of surface-tension driven spreading has long played an important role in our physical understanding of the dynamics of wetting~\cite{degennes-1985}. In surface-tension driven problems, on length scales smaller than the capillary length, $L_c = \sqrt{\gamma/\rho g}$, gravity is negligible (for surface energy $\gamma$ and fluid density $\rho$).  In this limit, an assumption that the thickness of droplet $h = 0$ at the contact line leads to divergent viscous stresses, and hence to the theoretical immobility of contact lines~\cite{huh-1971}.  This apparent paradox, which conflicts with everyday experience of spreading droplets, can be resolved by considering the development of a precursor film due to intermolecular interactions (van der Waals for example) in advance of the contact line~\cite{degennes-1985}. There, a local balance between viscous dissipation and the rate of change of surface energy gives rise to Tanner's law~\cite{bonn-2009,tanner-1979}, in which the droplet radius advances with speed $\rd R/\rd t\propto \theta^3$ for apparent contact angle $\theta$, and thus $R$ increases as $t^{1/10}$.

In the elastic case considered here, we show that while propagation is similarly controlled by dynamics at the peeling front, the dominant balance is now between viscous forces and elastic bending and tension.  The result is a rich set of solution behaviours in which spreading is governed by `peeling-by-bending' or `peeling-by-pulling' conditions at the peeling front.
\begin{figure}[h]
\includegraphics[width = 3in]{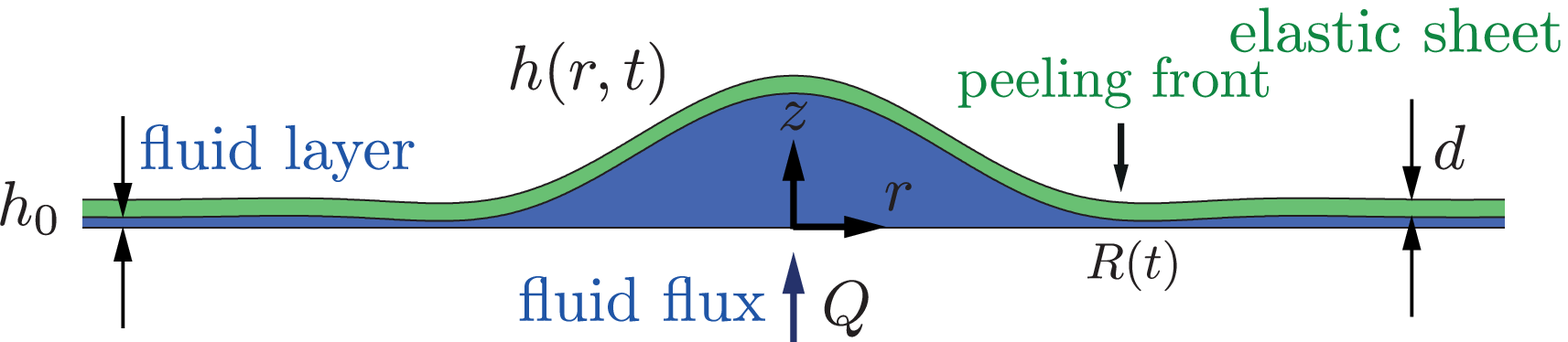} \\
\vspace{.05in}
\includegraphics[width = 3in]{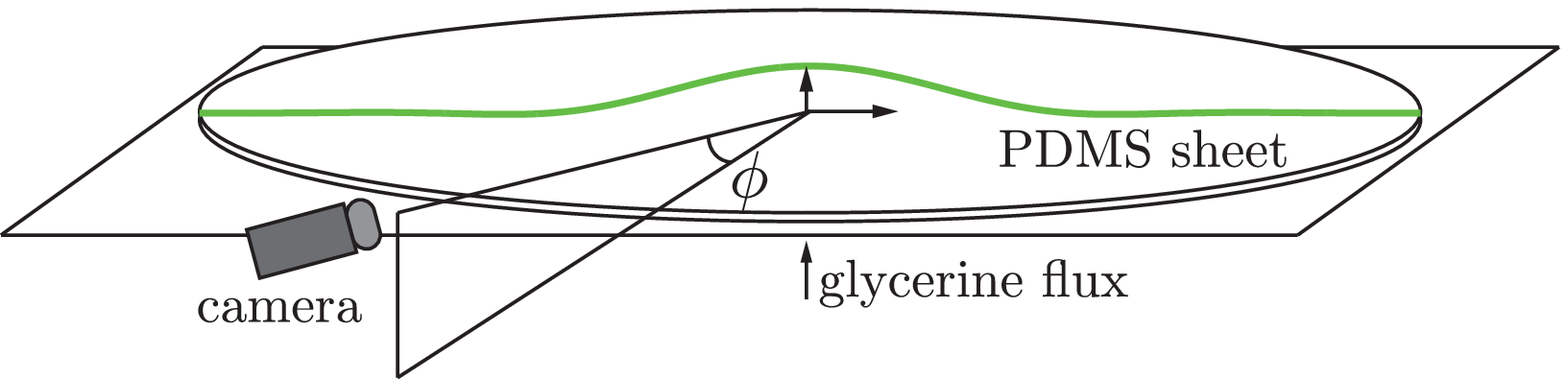}
\caption{Schematic of the model and experimental setup. \label{schematic}}
\end{figure}

We examine the peeling-by-bending regime theoretically and experimentally in the geometry illustrated in figure~\ref{schematic}. An axisymmetric fluid blister of thickness $h(r,t)$ is formed by a volumetric flux $Q$ of viscous fluid injected below an elastic sheet of thickness $d$ that rests on a thin pre-wetting layer of fluid of  thickness $h_0$ and viscosity $\mu$.  When the vertical deflection of the elastic sheet is small compared to its thickness, $h \ll d$,
we can neglect stretching of the sheet and consider only bending stresses.  The fluid pressure is the sum of elastic and hydrostatic components; in this case $p = B \nabla^4h + \rho g (h-z)$, where $B = Ed^3/12(1-\nu^2)$ is the bending stiffness of the sheet, and $E$ and $\nu$ are the Young's modulus and Poisson's ratio.  Lateral gradients in this pressure drive flow and thus, within the lubrication approximation, the evolution of the elastic sheet deflection and a global statement of mass conservation are given by~\cite{michaut-2011}
\begin{eqnarray}
\frac{\partial h}{\partial t} & = & \frac{1}{12\mu}
\boldsymbol\nabla\cdot
[h^3\boldsymbol\nabla(B\nabla^4h +  \rho g h)], \label{pde} \\
Qt & = & 2 \pi \int_0^{R(t)} (h-h_0)\, r\, \rd r, \label{vcons}
\end{eqnarray}

\noindent where $R(t)$ is the radial extent of the fluid blister. We can immediately identify in (\ref{pde}) a radial `elasto-gravity' length scale $\lb = (B/\rho g)^{1/4}$ at which bending stresses and gravity contribute equally to flow; $\lb$ is the analogue of the capillary length $L_c$.  Natural height and time scales for (\ref{pde}), (\ref{vcons}) are  $\lh = (12 Q\mu/\rho g)^{1/4}$ and $\tau = \lh \lb^2/Q$.  

As in the capillary case, gravity is negligible near a contact line, and any requirement that $h \rightarrow 0$ as $r\rightarrow R$ implies divergent viscous stresses or an immobile blister ($\rd R/\rd t = 0$)~\cite{flitton-2004,note3}.  In the presence of a pre-wetting layer, 
propagation must therefore be determined by matching to a solution for `peeling-by-bending' at the blister edge.
 
At early times, when $R \ll \lb$, we can neglect gravity and  consider a simpler peeling problem. For small pre-wetting film thicknesses, $\epsilon \equiv h_0/\lh \ll 1$, spreading is very slow and the interior has uniform pressure $p = B\nabla^4h$ with $h = h' = O(\epsilon)$ at $r = R$ and $h' = h''' = 0$ at $r = 0$. Thus the interior solution is
\begin{equation}
h(r,t) = \frac{3Qt}{\pi R^2(t)}\left(1-\frac{r^2}{R^2(t)}\right)^2. \label{interior}
\end{equation}

If $R$ is to increase, the elastic sheet near the blister edge must be peeled away from the pre-wetted substrate by bending.  A local travelling-wave solution with speed $c$ of the form $h = h_0 f(x-ct)$ must  satisfy 
\begin{equation}
-c f' = \frac{B h_0^3}{12 \mu}[f^3 (f'''')']'. \label{tip}
\end{equation}

We define a peeling length scale $\lp = (Bh_0^3/12\mu c)^{1/5}$, and solve $F^3F^{(v)} + F = 1$, where $F[\xi\equiv(x-ct)/\lp] = f(x-ct)$, subject to $F\rightarrow 1$ as $\xi\rightarrow\infty$ and $F''' = F'''' \rightarrow 0$ and $F'' \rightarrow A$ as $\xi\rightarrow-\infty$, in order to match to the curvature of the interior, constant-pressure, solution \cite{note3}.  Solving this system numerically, we find $A = 1.35$ and hence find the dimensional peeling speed  in terms of the curvature $\kappa$ of the interior solution at the tip. 
\begin{equation}
\frac{\rd R}{\rd t}=
c = \frac{B h_0^{1/2}}{12 \mu} \left(\frac{\kappa}{1.35}\right)^{5/2} \label{tipspeed}
\end{equation}

\noindent This new propagation law for peeling by bending is  the elastic analogue of Tanner's law~\cite{tanner-1979} for surface-tension driven spreading, and can be contrasted with previous solutions for inextensible peeling by pulling \cite{mcewan-1966}.

Using the new propagation law (\ref{tipspeed}), and the form of the interior solution in (\ref{interior}), we now find that the radius and height of the blister are given by similarity solutions
\begin{eqnarray}
R(t) & = & 1.31 \left(\frac{h_0 B^2 Q^5}{\mu^2}\right)^{1/22} t^{7/22},\label{R_bend} \\
h(0,t) & = & 0.55\left(\frac{\mu^2Q^6}{h_0 B^2}\right)^{2/22} t^{8/22},
\label{h_bend}
\end{eqnarray}

\noindent respectively.

We experimentally examined the dynamics of this peeling-by-bending solution by injecting a viscous fluid under a deformable elastic sheet and accurately measuring the surface deflections through time.  The experiments were performed using a $930\pm 2$ mm diameter PDMS sheet (Dow Corning Sylgard 184 silicone elastomer) with thickness  $d = 10 \pm 0.5\ \textrm{mm}$, Young's modulus $E = 1.82\pm0.09\ \mbox{MPa}$~\cite{schneider-2008}, Poisson's ratio $\nu = 0.45$, and therefore bending stiffness $B = 0.188\ \mbox{Pa m}^{3}$.  The PDMS sheet was placed on a rigid perspex base with a central 15.9 mm diameter hole through which fluid could be injected.  


Surface deflections of the PDMS sheet were measured by digitally imaging a pre-drawn line on the sheet from a known oblique angle $\phi$ to the horizontal (see figure~\ref{schematic}) and at right angles to the line.  For each experiment the deflection of the line was measured with respect to a reference image of the undeflected line.  Sub-pixel accuracy was achieved by fitting a gaussian profile across the line (whose width was $\sim 1.5\ \textrm{mm}$), processing the differences between the deflected and reference images, and thereby resolving vertical deflections of order 10$\,\mu$m \cite{note3}. 

Experiments were prepared by injecting a known small volume of glycerine under the PDMS sheet, and manually spreading the fluid evenly over the full area of the sheet.  This provided an estimate of the average pre-wetting film thickness $h_0$. A local measure of the pre-wetting film thickness was provided by observing deflection of the sheet by a small weight.  During the experiment glycerine was injected under the centre of the PDMS sheet with a peristaltic pump (Watson-Marlow 502s) and the mass flux measured with a digital scale (Ohaus Adventure Pro)~\cite{note1}. 

\begin{figure}
\includegraphics[width = 3.3in]{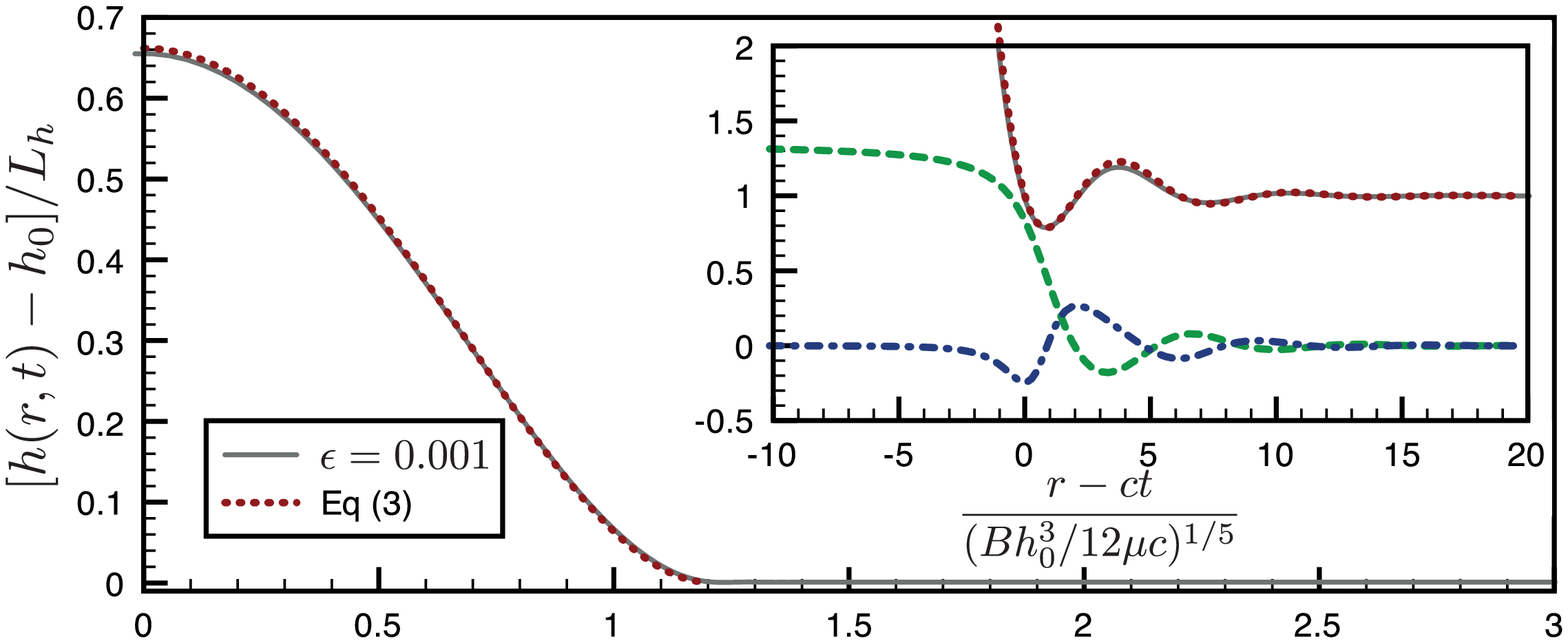}
\includegraphics[width = 3.3in]{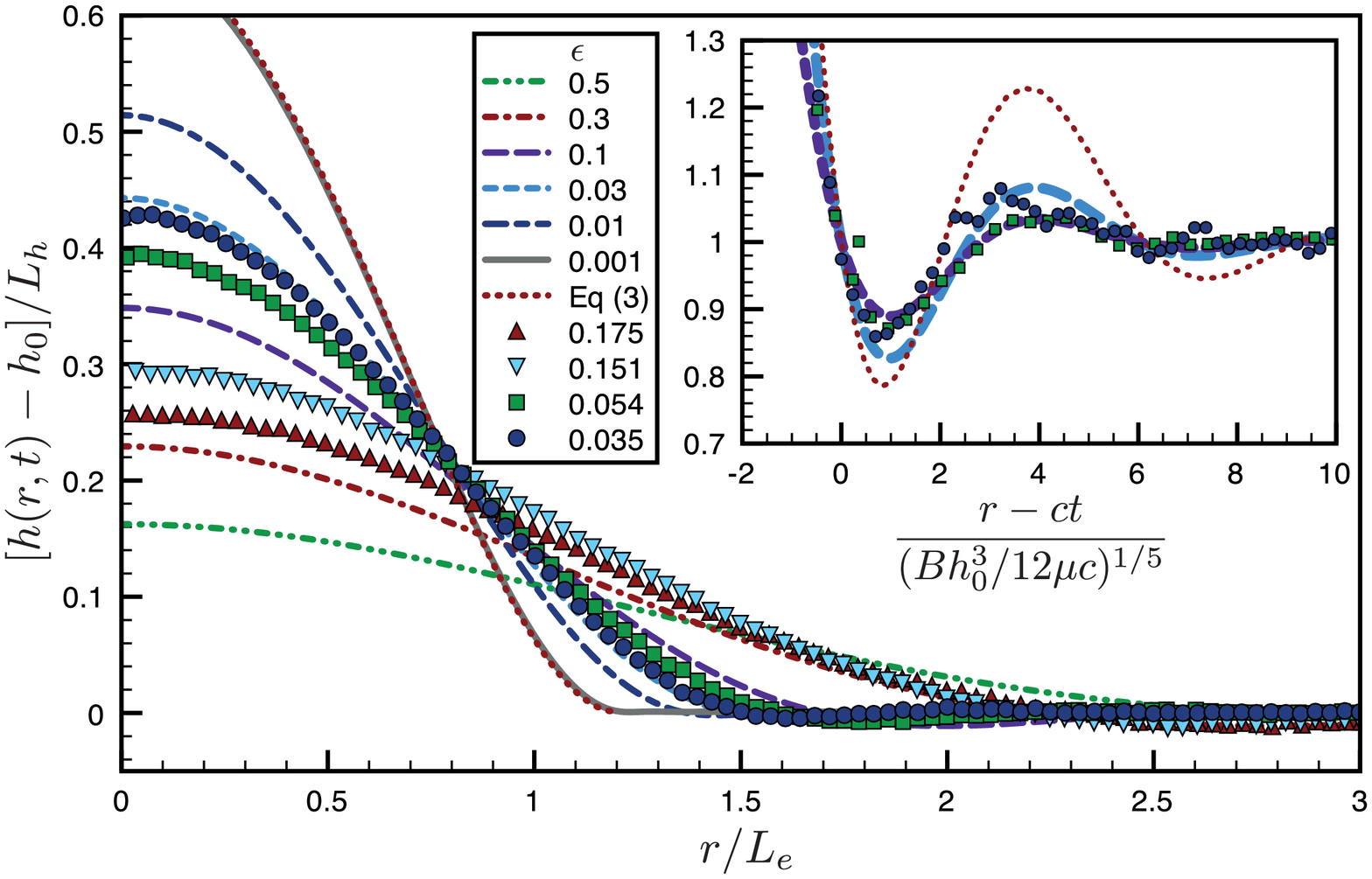}
\caption{Asymptotic, numerical and experimental profiles of the elastic blister at $t/\tau = 1$. (a) Asymptotic and numerical solutions for $\epsilon\equiv h_0/\lh = 0.001$. Inset shows the peeling-by-bending travelling-wave solution $F$ to (\ref{tip}). (b) Experimental profiles for $\epsilon = 0.035,$ 0.054, 0.151, and 0.175, with other experimental parameters detailed in~\cite{note1}, and numerical solutions for $\epsilon= 0.5,\dots,0.001$.   Inset shows the scaled experimental profiles at the blister edge for $\epsilon = 0.035$ and 0.054 with numerical solutions for $\epsilon =0.03$ and 0.1.  \label{profiles}}
\end{figure}
\begin{figure}
\includegraphics[width = 3in]{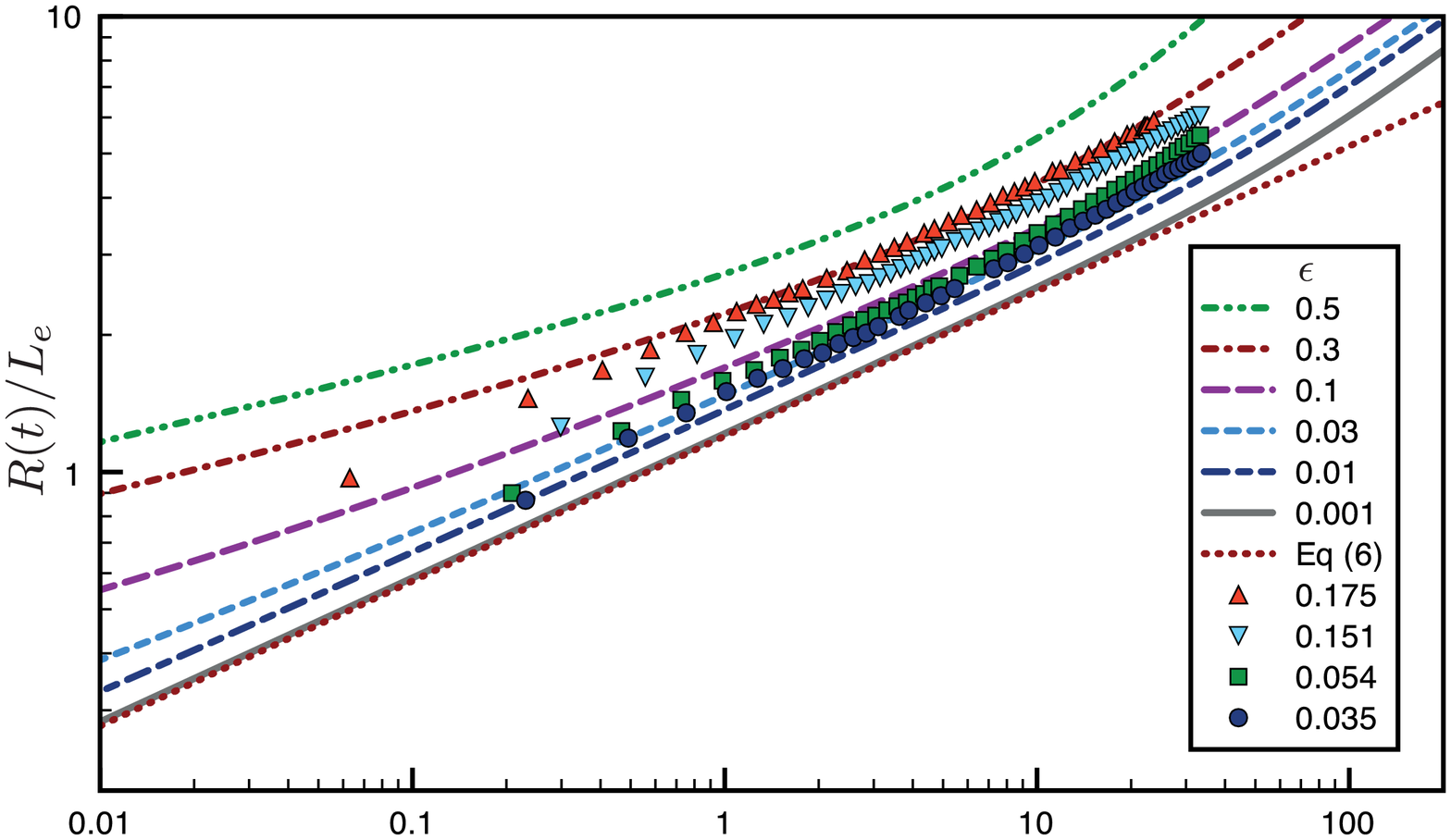}
\includegraphics[width = 3in]{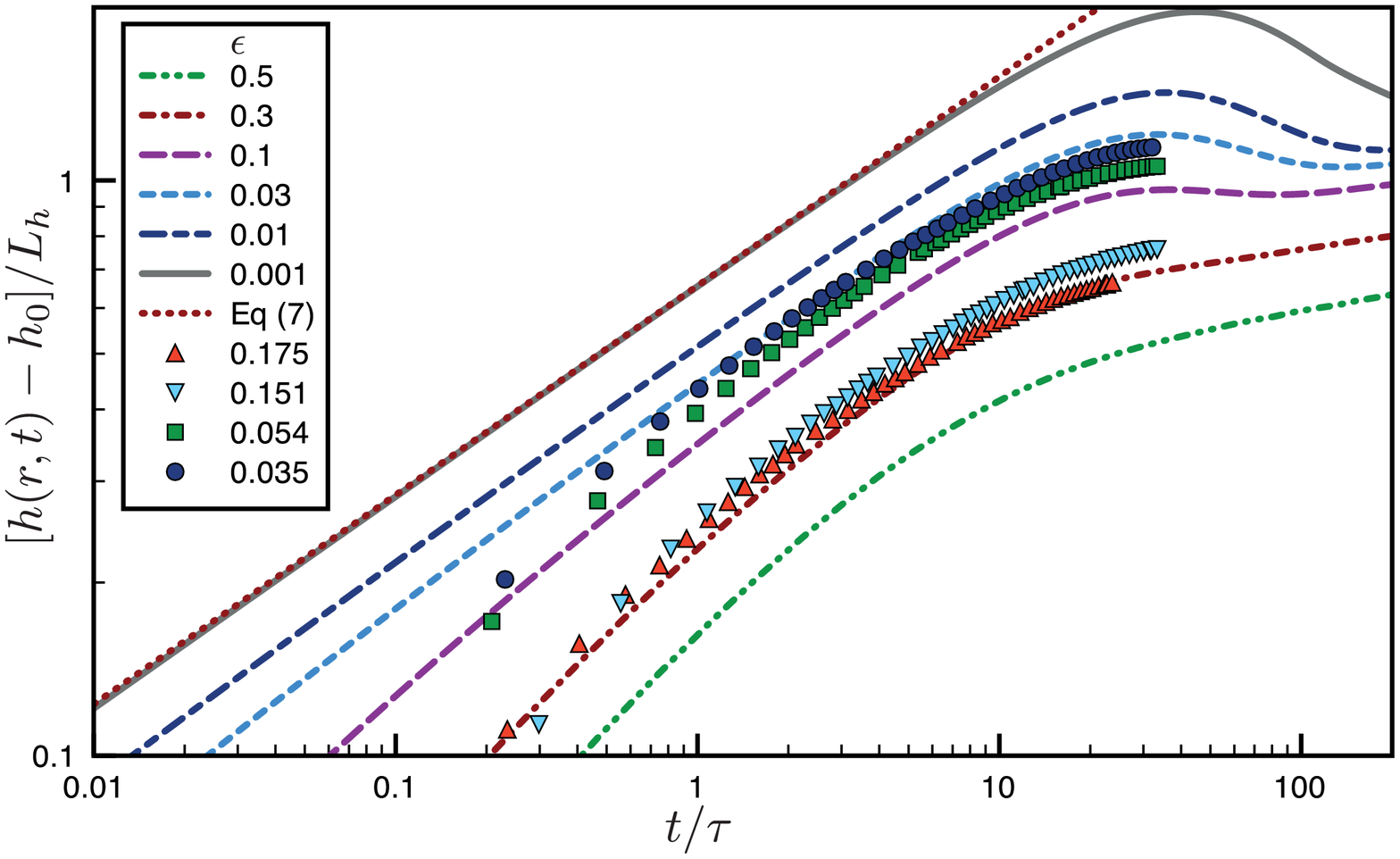}
\caption{Numerical and experimental results, with similarity solutions (\ref{R_bend}) and (\ref{h_bend}) for $\epsilon=0.001$. (a) Dimensionless radius with time.  (b) Dimensionless height at the origin with time. \label{RH} }
\end{figure}

The results of these experiments are shown in figures~\ref{profiles}b and~\ref{RH}, with comparisons to numerical solutions of the evolution equation (\ref{pde}) for various pre-wetting film thicknesses $h_0$.  The data, scaled using the elasto-gravity length $\lb$ and time scale $\tau$ (see~\cite{note1}), confirm 
that the radial extent is a function of the pre-wetting film thickness $h_0$ and thus demonstrate the importance of edge control by peeling.  The inset shows the comparison between the experimental profiles for $\epsilon = 0.035$ and 0.054 with the numerical solutions of (\ref{pde}) and (\ref{tip}) for the peeling-by-bending wave.  We can see evidence for the flexural wave, with a dimensional amplitude of about $30\,\mu \textrm{m}$.  

Figure~\ref{RH} shows the radial extent  (top) and central height of the blister  (bottom) as functions of the scaled time. For $\epsilon \ll1$ and $R\ll \lb$ there is excellent agreement with the similarity solutions (\ref{R_bend}) and (\ref{h_bend}) \cite{note4}. For $R\gg \lb$ there is a clear transition to a new regime because gravity can no longer be neglected in the form of the interior solution.

At intermediate times, when $\epsilon\ll 1$ and $R\gg \lb$, peeling by bending continues to control propagation but gravity now plays an increasing role in the interior. The interior blister remains quasistatic with negligible horizontal pressure gradients and satisfies $\lb^4\nabla^4 h+h=\textrm{constant}$ with $h(R)=h'(R)=0$ and $h' = h''' = 0$ at $r=0$.  Solutions have the asymptotic form of an interior flat-topped region, of height $h_i$, with a peripheral bending region of width $O(\lb)$, where  
\begin{equation}
h(y)=h_i[1-{\rm e}^{-y}(\cos y+\sin y)]
\label{edge_grav}
\end{equation}
and $y=(R-r)/\sqrt2 \lb$. This is the elastic analogue of a sessile drop  at large Bond number, which forms a flat-topped puddle with a peripheral meniscus on the scale of the capillary length~\cite{hocking-1983}.  

The interior curvature of (\ref{edge_grav}) at the peeling front, $r=R$, is now $\kappa=h_i/\lb^2$, and the volume constraint (\ref{vcons}) gives $h_i\approx Qt/(\pi R^2)$. The elastic spreading law (\ref{tipspeed}) thus implies
\begin{eqnarray}
R(t) & = & 0.40 \left(\frac{\rho g}{B}\right)^{5/24}
\left(\frac{h_0 B^2 Q^5}{\mu^2}\right)^{1/12} t^{7/12}, 
\label{R_grav}\\
h(0,t) & = & 2.02\left(\frac{B}{\rho g}\right)^{5/12}
\left(\frac{\mu^2Q}{h_0 B^2}\right)^{1/6} t^{-1/6}.
\label{h_grav}
\end{eqnarray}
Note that the height is predicted to decrease in this regime, explaining the numerical behaviour seen in figure \ref{RH} for $\epsilon \leqslant 0.03$.

At late times, where $R\gg \lb \epsilon^{-1/2}$, the pressure drop associated with the interior Poiseuille flow from the source towards the peeling front becomes the dominant resistance to propagation. The flow enters a new regime in which the bending stresses in (\ref{pde}) can be neglected almost everywhere, resulting in a standard viscous gravity-current balance~\cite{michaut-2011,huppert-1982}. The extent
\begin{equation}
R(t)=0.715\ (\rho gQ^3/12\mu)^{1/8}t^{1/2}, \label{R_gc}
\end{equation}
and while bending stresses modify the shape of the gravity-current solution near $r=0$ and $r=R$, they no longer control the dynamics of propagation.

In summary, for $h_0\ll h\ll d$ the flow passes through three asymptotic dynamical regimes, as confirmed numerically: 
pressure-driven peeling with $R(t)$ given by (\ref{R_bend}) for $R\ll\lb$ (or $t/\tau\ll \epsilon^{-1/7}$); 
gravity-driven peeling given by (\ref{R_grav}) for $1\ll R/\lb \ll  \epsilon^{-1/2}$ (or $\epsilon^{-1/7}\ll t/\tau\ll \epsilon^{-1}$); 
and a viscous gravity current given by (\ref{R_gc}) for $R\gg \lb \epsilon^{-1/2}$. Our experiments straddle the first two of these regimes.

A different analysis is required when the deflection $h(r,t)$ of the elastic sheet is large compared to its thickness $d$. At large $Q$ this could happen even if $h_0\ll d$ through the $t^{8/22}$ growth in (\ref{h_bend}) before any transition to (\ref{h_grav}). In the experiments of~\cite{puzovic-2012}, $d$ and $h_0$ were both in the range 0.33--0.97\,mm, and the thinness of their latex sheets meant $h(0,t)/d$ reached values of order 10. In these circumstances, the stretching of the sheet can no longer be neglected when calculating the elastic stresses and fluid pressure. 

The F\"oppl-von-Karman plate equations for an axisymmetric pressurised blister~\cite{jensen-1991} can be written as
\begin{eqnarray}
p=B\nabla^4 h - \frac{1}{r} \frac{\rd}{\rd r} \left( r T \frac{\rd h}{\rd r}\right),\label{FK1}\\
\frac{1}{r} \frac{\rd}{\rd r}  \left( r^3 \frac{\rd T}{\rd r}\right) =-\frac{Ed}{2}\left(\frac{\rd h}{\rd r}\right)^2,\label{FK2}
\end{eqnarray}
where $T(r,t)$ is the radial tension in the sheet induced by  stretching. Scaling shows that for $h\ll d$ the tension term in (\ref{FK1}) can be neglected, thus recovering (\ref{pde}). Conversely, for $h\gg d$ the bending term in (\ref{FK1}) can be neglected in the interior. 

Assuming that a slow peeling process controls the rate of spread, we again expect a constant-pressure interior solution for $h\gg d$. After integration of (\ref{FK1}) to find $\rd h/\rd r=-rp/2T$ in $r<R$, (\ref{FK2}) yields
\begin{equation}
\frac{T^2}{r^3} \frac{\rd}{\rd r}  \left( r^3 \frac{\rd T}{\rd r}\right) =-\frac{Ed\,p^2}{8}.\label{FK3}
\end{equation}
In $r>R$ equation (\ref{FK2}) yields $T\propto r^{-2}$.
We solved (\ref{FK3}) numerically subject to regularity at $r=0$ and the matching condition $(r^2T)'=0$ at $r=R$. The solution describes the tension and hence shape of the stretched sheet. The volume constraint (\ref{vcons}) gives the fluid pressure as $p=0.324\,Ed(Qt)^3/R^{10}$, and the sheet approaches $r=R$ with a contact angle $\theta=1.64\,Qt/R^3$ and edge tension $T_\theta=0.099\,Ed(Qt/R^3)^2$. (The tension at $r=0$ is $1.71T_\theta$.) This solution is the elastic analogue of the spherical-cap shape~\cite{tanner-1979} of a capillary drop with a small contact angle.

There are two possibilities for the rate of spread of the pressurised elastic blister, depending on the relative sizes of the peeling length scale $\lp$ and a bending boundary-layer length scale $\lbl=(B/T_\theta)^{1/2}$ that arises from a balance of the two terms in (\ref{FK1}) near $r=R$:

If $h_0\ll d$ then there is a static bending boundary layer, where 
\begin{equation}
h'=\theta({\rm e}^{(r-R)/\lbl}-1),\label{bbl}
\end{equation}
within which is nested a peeling-by-bending travelling-wave solution of the form analysed in the first part of the paper. Evaluating the curvature $\kappa$ from (\ref{bbl}) and using the propagation law (\ref{tipspeed}), we deduce that
\begin{equation}
R(t)=0.783 \left(\frac{Ed h_0^2}{B}\right)^{5/64}\left(\frac{BQ^5}{\mu h_0^2}\right)^{1/16}t^{3/8}
\label{R_membend}
\end{equation}

Alternatively, if $h_0\gtrsim d$ then a bending boundary layer is unnecessary since the viscous pressure drop of the peeling wave extends over a length scale $h_0/\theta$ greater than $\lbl$. Peeling is then by pulling with tension $T_\theta$, locally like an inextensible tape~\cite{mcewan-1966}.
Matching the interior solution (\ref{FK3}) to the Landau--Levich peeling-by-pulling solution~\cite{bonn-2009} yields a propagation law of Cox--Voinov type,
\begin{equation}
\frac{\rd R}{\rd t}=\frac{T_\theta\theta^3 }{36\mu\ln(1/\delta)},
\label{CV}
\end{equation}
where $\delta$ is the ratio of inner and outer length scales. (The factor 36, rather than 9 in capillary wetting, arises from the no-slip condition at the sheet.) Combining (\ref{CV}) with the numerical solutions for $T_\theta$ and $\theta$ gives
\begin{equation}
R(t)=0.807\left(\frac{EdQ^5}{\mu \ln(1/\delta)}\right)^{1/16}t^{3/8},
\label{R_mempull} 
\end{equation}
where, for simplicity, we take $\delta=h_0/\theta R$. (An alternative theory, with which we disagree \cite{note3}, is given in \cite{al-housseiny-2013}.)

\begin{figure}
\includegraphics[width = 3.3in]{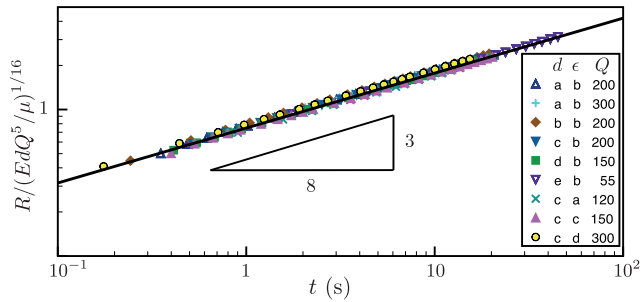}
\caption{Collapse of experimental data from a Hele-Shaw cell with an elastic wall~\cite{puzovic-2012}. The raw data is the same as in  their fig.~2a, and is replotted with approximately corresponding symbols for a range of flow rates $Q\ [\mbox{cm$^3$min}^{-1}]$, sheet thickness $d$ (their $h$), and pre-wetting film thickness $h_0$; $h_0/d$ varies from 0.57 to 1.7. See~\cite{puzovic-2012} for details. An average value $1/\delta 
\simeq 30$ was used when evaluating the line $0.748t^{3/8}$ from (\ref{R_mempull}).\label{draga}}
\end{figure}

In figure \ref{draga} we compare the experimental data of~\cite{puzovic-2012} with the theoretical prediction (\ref{R_mempull}). We note that there is significantly better collapse of the data than in figure 2b of~\cite{puzovic-2012}, where the scaling differed by a factor $(h_0/d)^{1/8}$~\cite{note2}, and that there is excellent agreement with the theory. 

This agreement might initially be thought surprising since spread in~\cite{puzovic-2012} was driven by gas rather than fluid injection. However, if the pre-wetting fluid accumulates in the peeling wedge, a simple volume balance shows that its radial extent $x\sim(h_0\theta/R)^{1/2}\propto t^{2/8}$ is greater than the scale $h_0/\theta\propto t^{1/8}$ of the peeling region. Thus the gas is irrelevant to the predicted rate of spread (except perhaps by about 3\% if we instead take $\delta=h_0/\theta x$).

Late-time suppression of Saffman--Taylor fingering in~\cite{puzovic-2012} can be explained by the decrease in the capillary number~\cite{al-housseiny-2012}, but not, on its own, the complete suppression of instability for small $Q$. We hope that our theoretical solution for the radial base state will shed light on the instability mechanism. More importantly, we have shown here that elastic peeling away from a pre-wetting film is the dominant control on propagation in a suite of problems. Peeling-by-bending according to (\ref{tipspeed}) is a novel variation on peeling-by-pulling at the tip (\ref{CV}).


\begin{acknowledgments}
We thank D. Vella for many valuable discussions about these problems.
M.A. Hallworth assisted with the experiments. J.A.N. is supported by a Royal Society University Research Fellowship. 
\end{acknowledgments}

\bibliography{lacc9}

\begin{thebibliography}{23}
\bibitem [{Note1()}]{note1}%
  \BibitemOpen
  \bibinfo {note} {Experimental parameters: With $\rho =
  1.26\,\mbox{g\,cm$^{-3}$}$ and $B = 0.188\,\mbox{Pa\,m$^2$}$, the radial
  scale $\lb = 62.5\,\mbox{mm}$. For $\epsilon\equiv h_0/\lh =
  (0.035,0.054,0.151,0.175)$, respectively, $Q =
  (2.493,2.475,2.448,1.553)\,\mbox{cm$^3$s$^{-1}$}$, $\nu =
  (9.52,9.68,9.52,12.3)\,\mbox{cm$^2$s$^{-1}$}$ and $h_0 = (260,$
  $400,1110,1220)\,\mu\mbox{m}$, while the height and time scales are $\lh =
  (7.3,7.4,7.3,7.0)\,\mbox{mm}$ and $\tau =$
  $(11.5,11.6,11.6,17.5)\,\mbox{s}$.}\BibitemShut {Stop}%
\bibitem [{Note4()}]{note4}%
  \BibitemOpen
  \bibinfo {note} {For larger $\epsilon$ the early behaviour of (1) and (2) has
  $h\approx h_0$, and $R\propto t^{1/6}$ and $h-h_0\propto
  t^{2/3}$.}\BibitemShut {Stop}%
\end{thebibliography}

\end{document}